\documentclass[sigconf,nonacm]{acmart}
\begin{document}
\title{Uncertainty Quantification for Financial Foundation Models: A Survey}
\titlenote{Accepted at the 7th ACM International Conference on AI in Finance (ICAIF 2026). ACM DOI: \url{https://doi.org/10.1145/3858063.3858824}. This author version is distributed under the Creative Commons Attribution 4.0 International License (CC BY 4.0).}

\author{Zijie Zhao}
\orcid{0009-0006-1835-6197}
\affiliation{
  \institution{Massachusetts Institute of Technology}
  \city{Cambridge}
  \state{MA}
  \country{USA}
}
\email{zijiezha@mit.edu}

\author{Mingjun Sun}
\orcid{0009-0008-9983-3294}
\affiliation{
  \institution{Carnegie Mellon University}
  \city{Pittsburgh}
  \state{PA}
  \country{USA}
}
\email{mingjuns@andrew.cmu.edu}

\author{Shenbo Xu}
\orcid{0000-0002-4919-1034}
\affiliation{
  \institution{Massachusetts Institute of Technology}
  \city{Cambridge}
  \state{MA}
  \country{USA}
}
\email{xushenbo@mit.edu}

\begin{abstract}
Financial foundation models increasingly support workflows spanning financial question answering, disclosure analysis, forecasting, and decision-making. In these settings, uncertainty may concern whether evidence is valid and available at the relevant time, whether an interpretation is justified, which outcomes remain plausible, or which action is preferred under costs and risk constraints. Because these objects arise at different points in a workflow, uncertainty cannot be reduced to confidence in a final output. We introduce a source-aware framework that distinguishes six sources of uncertainty across evidence and context, model and inference, and outcome and action, while treating agentic workflows as the execution layer through which uncertainty can propagate, change form, or be reduced. We develop a mechanism-based taxonomy grounded in how methods construct or verify uncertainty and reliability representations, and use it to assess their applicability across financial task domains. We further map finance-specific empirical studies across the taxonomy and task domains to characterize the current research landscape. Our analysis shows that most reviewed studies address uncertainty in individual parts of financial workflows, with limited work connecting evidence and reasoning reliability to downstream prediction and constrained action. By connecting where uncertainty arises, how it is represented, and how it bears on downstream decisions, this survey provides a unified foundation for studying and evaluating uncertainty in financial foundation models.
\end{abstract}


\maketitle

\section{Introduction}
\label{sec:intro}

Financial foundation models increasingly support workflows that connect financial information to analysis, forecasting, and decisions. Their outputs may be fluent and plausible yet unsupported, numerically inconsistent, or based on information unavailable at the relevant decision time. Uncertainty quantification (UQ) can help identify unreliable cases, but uncertainty in financial workflows is not confined to the final output. Consider a system that retrieves an earnings release, interprets revised guidance, forecasts the market response, and recommends a trade. At each stage, a different question may remain unresolved: whether the evidence was available and relevant at the decision time, whether the interpretation follows from the reported figures and market context, which outcomes remain plausible, and which feasible action is preferred under stated objectives, costs, and risk constraints. Because these questions concern distinct uncertain objects, uncertainty cannot be reduced to confidence in the final output: such a score cannot indicate whether the system should retrieve additional evidence, verify its reasoning, revise its forecast, or change its action. Financial UQ must therefore identify what is uncertain, represent the remaining uncertainty, and determine whether system behavior should change.

Throughout this survey, \emph{uncertainty} denotes the extent to which a quantity, proposition, outcome, or preferred action remains unresolved relative to a specified information set and modeling assumptions; for decisions, it is also relative to stated objectives and constraints. Its representation depends on the object and may take the form of a support judgment, confidence score, disagreement measure, predictive distribution, interval, or prediction set. \emph{Confidence} is a score attached to a prediction, answer, or claim and need not be calibrated. \emph{Calibration} describes the correspondence between stated probabilities or confidence levels and empirical frequencies over comparable cases~\citep{guo2017calibration}. We use \emph{aleatoric uncertainty} for variability that remains conditional on the specified information set and assumptions, and \emph{epistemic uncertainty} for uncertainty arising from incomplete evidence, data, knowledge, or model specification that is reducible in principle. The object-based framework identifies what remains unresolved, whereas the aleatoric--epistemic distinction characterizes the nature of that uncertainty; each object may involve both components.

General LLM UQ surveys synthesize work on confidence estimation, semantic consistency, calibration, conformal methods, and abstention~\citep{shorinwa2025survey,he2025sources}. Financial UQ surveys focus primarily on probabilistic modeling and uncertainty estimation for financial time series~\citep{blasco2024financialuq,eggen2026financialuq}. Surveys of financial language models and agents instead emphasize models, tasks, agent architectures, and deployment settings~\citep{li2023finance_survey,dong2025financeagents}. These lines of work provide important foundations but do not jointly organize point-in-time evidence validity, task-level capability, case-specific reasoning and interpretation, uncertain financial outcomes, and actions under stated objectives and constraints, or explain how these forms of uncertainty interact across retrieval, reasoning, tool use, memory, verification, and execution. We address this intersection by organizing uncertainty around the object that remains unresolved and tracing its role across financial foundation-model workflows, including agentic systems.

\textbf{Survey scope and methodology.} We focus on uncertainty in financial workflows supported by language, multimodal, and time-series foundation models, while drawing on established general and financial UQ where they provide relevant methodological or evaluative foundations. We searched ACM Digital Library, ACL Anthology, IEEE Xplore, arXiv, SSRN, and major machine-learning and finance venues for work available through July 25, 2026. The search combined terms for financial language, time-series, and agentic foundation models with uncertainty, calibration, grounding, verification, probabilistic forecasting, risk, and distribution shift. We included finance-specific studies that construct, verify, propagate, evaluate, or use operational uncertainty or reliability representations, together with directly relevant methodological and evaluation foundations. Duplicate preprint, working-paper, and published versions were merged. Section~\ref{sec:landscape} defines the narrower empirical subset used for the coverage analysis.

We make three contributions:
\begin{itemize}
  \setlength{\itemsep}{1pt}
  \setlength{\topsep}{2pt}
  \setlength{\parsep}{0pt}
  \setlength{\parskip}{0pt}
  \item \textbf{A source-aware framework.} We distinguish six uncertainty sources across evidence and context, model and inference, and outcome and action, and treat agentic workflows as the execution layer through which these sources interact and propagate.
  \item \textbf{A mechanism-based taxonomy and task analysis.} We classify methods by how they construct or verify uncertainty and reliability representations and analyze their applicability across financial task domains.
  \item \textbf{An empirical landscape and research priorities.} We map finance-specific empirical studies across method families and task domains to identify concentrations, missing connections, and research priorities.
\end{itemize}

\section{A Source-Aware Framework for Financial Uncertainty}
\label{sec:taxonomy}

\begin{figure*}[t]
  \centering
  \includegraphics[width=\textwidth]{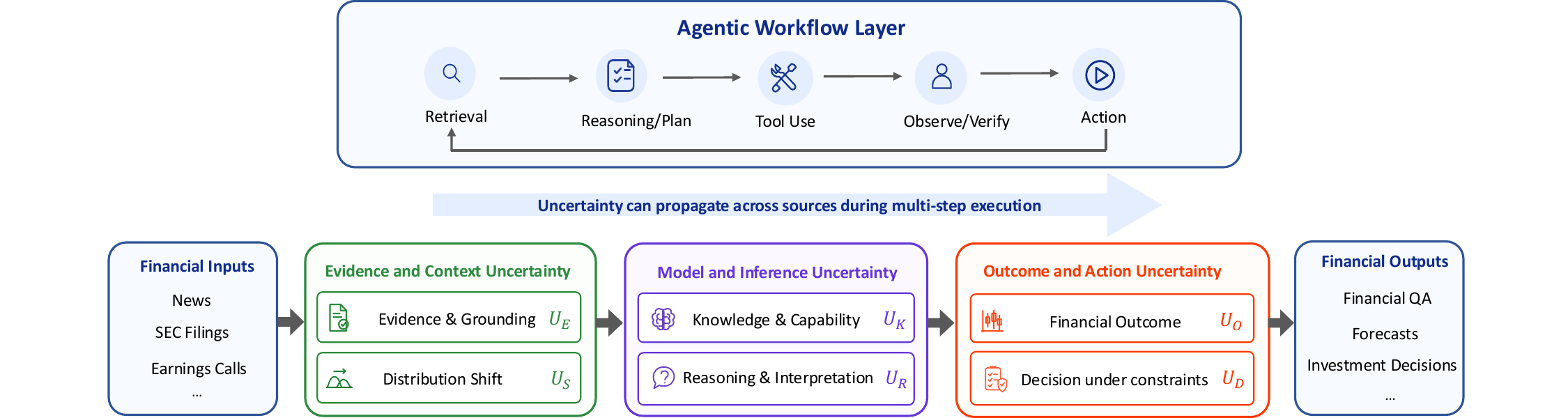}
  \Description{Six sources of financial uncertainty are organized into evidence and context, model and inference, and outcome and action. An agentic workflow layer spans retrieval, reasoning and planning, tool use, observation and verification, and action, with feedback across iterations.}
  \caption{A source-aware framework for financial uncertainty. The agentic workflow layer connects the six sources as uncertainty propagates, changes form, or is reduced during execution.}
  \label{fig:framework}
\end{figure*}

We propose a source-aware framework that organizes financial uncertainty by the object that remains unresolved in a foundation-model workflow, rather than by a fixed pipeline stage or downstream failure. Figure~\ref{fig:framework} summarizes six sources across evidence and context, model and inference, and outcome and action, together with the agentic execution layer that connects them. Several sources may coexist in the same case; when a primary source is needed, we assign it according to what is directly uncertain.

\subsection{Sources of Financial Uncertainty}

\textbf{Evidence and context uncertainty} concerns the validity of the available information and whether previously established reliability remains applicable. \textbf{Evidence and grounding uncertainty} (\(U_E\)) asks whether a claim is supported by evidence that is correct, sufficiently complete, traceable, and available at the relevant decision time. In finance, an authentic filing cannot support a historical decision if it was released or restated only afterward, while memorized future information can contaminate retrospective analysis~\citep{kelly2026scalingpit,lopez2025memorization}. \textbf{Distribution shift uncertainty} (\(U_S\)) asks whether reliability established during training or calibration transfers to the current environment as market regimes, disclosure practices, accounting standards, or the population of firms and instruments changes~\citep{ovadia2019trust,gibbs2021adaptive}. Thus, \(U_E\) concerns the validity of evidence in the current case, whereas \(U_S\) concerns whether established reliability transfers to the current setting.

\textbf{Model and inference uncertainty} distinguishes whether a system has the knowledge and capability required for a class of tasks from whether its reasoning and interpretation are adequate in the current case. \textbf{Knowledge and capability uncertainty} (\(U_K\)) asks whether the system has the domain knowledge and skills required for a class of financial inputs, including accounting conventions, numerical reasoning over filings, structured-data understanding, and tool use~\citep{kang2023hallucination,xie2024finben}. \textbf{Reasoning and interpretation uncertainty} (\(U_R\)) concerns whether the model's inference from the available evidence and stated assumptions to a specific conclusion is logically, numerically, and economically justified. It may arise from ambiguity in the problem formulation, faulty decomposition, numerical or tool-use errors, inappropriate comparison periods, omitted premises, or unsupported economic assumptions~\citep{dai2026realfin,yao2026fedspeak}. Because internal reasoning is not always observable, \(U_R\) is assessed through available intermediate outputs, calculations, tool calls, stated assumptions, and the consistency of the conclusion with the evidence.

\textbf{Outcome and action uncertainty} separates uncertainty about future financial results from uncertainty about how to act on them. \textbf{Financial outcome uncertainty} (\(U_O\)) concerns the conditional distribution of a defined financial target and horizon, such as returns, earnings, volatility, default risk, or portfolio loss~\citep{goel2024var,pele2026llmvar}. This uncertainty does not disappear when the evidence and interpretation are sound. Market outcomes remain uncertain because prices reflect prior expectations and strategic responses, while subsequent information, liquidity conditions, and future shocks affect the realized result~\citep{mackinlay1997event}. \textbf{Decision uncertainty under constraints} (\(U_D\)) concerns the preferred feasible action and its stability under uncertainty in outcomes, objectives, costs, liquidity, risk limits, and policy requirements. Known constraints define the feasible set; uncertainty lies in which feasible action is preferred and how sensitive that choice is to its inputs~\citep{sanderink2026alpha,jensen2026implementable}. Thus, \(U_O\) concerns what may happen, whereas \(U_D\) concerns what should be done.

The aleatoric--epistemic distinction complements rather than replaces the source framework. Evidence, capability, reasoning, and distribution-shift uncertainty often contain substantial epistemic components when additional information, data, computation, verification, or model improvement can reduce them. Reasoning and interpretation uncertainty may also contain aleatoric components when the task or evidence admits multiple defensible interpretations. Financial outcome uncertainty is mixed: uncertainty in the estimated conditional distribution may be reducible, while part of realized variation remains aleatoric under the stated information set and assumptions. Decision uncertainty inherits both forms through uncertain outcomes and model inputs. The distinction is therefore conditional on the information set and modeling assumptions and, for decisions, the stated objectives and constraints; it is not an intrinsic label attached permanently to a source.

\subsection{Agentic Workflow Layer and Uncertainty Propagation}
\label{sec:agentic_layer}

An agentic workflow is not a seventh source but an execution layer connecting the six sources across retrieval, reasoning and planning, tool use, memory, verification, and action. Intermediate outputs become inputs: evidence shapes interpretation, interpretation informs outcome estimates, and those estimates guide actions. A filing unavailable at the decision time may support an apparently coherent interpretation and subsequently distort prediction and action. Actions can change later observations, stored state, and available choices, making uncertainty trajectory-dependent~\citep{zhao2025saup,oh2026agentuq}.

Propagation need not be monotonic or irreversible. Retrieval, recomputation, provenance checks, or numerical verification may reduce epistemic uncertainty, although later steps may leave it unresolved or introduce uncertainty. When uncertainty cannot be resolved before action, abstention or escalation can limit its consequences rather than remove it. The source framework identifies what remains uncertain; the agentic layer describes how uncertainty propagates, changes form, or is mitigated across steps.

\section{Methods and Evaluation for Financial Uncertainty Quantification}
\label{sec:methods}

We focus on UQ methods used in or directly applicable to financial workflows supported by foundation models, drawing on established work on uncertainty representation and evaluation. We group these methods into five families based on how they construct or verify uncertainty or reliability representations: assessment against supporting evidence, direct estimation from a model, repeated execution of a fixed system, comparison across model or parameter realizations, or probabilistic modeling of a defined target. This taxonomy distinguishes methods by how their representations are obtained rather than by application, mathematical tradition, or downstream use. For hybrid methods, the primary family is determined by the mechanism producing the main uncertainty result.

Bayesian modeling may produce parameter variation, predictive distributions, or sequential belief updates, while calibration evaluates a representation rather than defining how it was constructed. Abstention, deferral, and action adjustment use uncertainty downstream and are therefore not separate method families.

\subsection{Methods for Constructing Uncertainty Representations}
\label{subsec:characterization}

\paragraph{Evidence and provenance verification}
These methods assess a claim or result against supporting evidence and provenance information valid at the relevant decision time. Attribution methods align claims with cited passages, while financial applications must also check document provenance, version, numerical consistency, and point-in-time availability~\citep{gao2023alce,kelly2026scalingpit,lopez2025memorization}. Recent methods decompose financial answers into atomic claims, verify them against retrieved evidence, and detect or revise unsupported statements~\citep{guo2026finground,tan2025fred,yin2026rlfkv}. Finance-specific benchmarks cover long-document claim verification, hallucination detection in grounded financial QA, and justified abstention in multi-turn financial RAG~\citep{zhao2024findver,kumar2026finreflectkg,cho2026kfinhallu}. We include verification in this taxonomy only when it produces a claim- or instance-level judgment of support or validity; aggregate retrieval or task accuracy alone is not an uncertainty representation. This family primarily characterizes evidence and grounding uncertainty (\(U_E\)) and may also expose unsupported reasoning (\(U_R\)).

\paragraph{Direct uncertainty estimation}
Direct methods produce an instance-level estimate without relying on repeated stochastic executions or multiple model realizations. They use token probabilities or entropy, verbalized confidence, self-evaluation, hidden-state features, or auxiliary error predictors~\citep{kadavath2022language,tian2023just,lahlou2023deup}. Financial applications estimate confidence in monetary-policy interpretation, identify incorrect answers from model representations, and predict epistemic error in cross-sectional stock ranking~\citep{yao2026fedspeak,wang2026confidentlywrong,sanderink2026alpha}. Probability-based estimates require access to model probabilities, whereas learned probes may require internal states and labeled data; neither is necessarily calibrated or aligned with downstream decision uncertainty. This family mainly characterizes knowledge and capability uncertainty (\(U_K\)) and reasoning and interpretation uncertainty (\(U_R\)).

\paragraph{Sampling-based consistency estimation}
These methods keep the underlying model or system fixed and compare outputs across stochastic generations, input perturbations, or repeated executions. SelfCheckGPT measures agreement across sampled generations, while semantic uncertainty methods group meaning-equivalent answers before measuring variation~\citep{manakul2023selfcheckgpt,kuhn2023semantic,farquhar2024semantic}. A finance-specific hybrid combines multi-sample semantic entropy with token-level and evidence-use features~\citep{singha2025eclipse}. Agreement does not establish correctness because repeated outputs may share the same error, and repeated execution adds latency and cost. Variation across runs mainly characterizes instability in \(U_K\) and \(U_R\); under input perturbations, it may also reveal sensitivity to distribution shift (\(U_S\)).

\paragraph{Ensemble-based uncertainty estimation}
Ensemble methods compare predictions from multiple models or parameter realizations. Deep ensembles use independently trained models, while Monte Carlo dropout approximates parameter sampling through repeated forward passes~\citep{gal2016dropout,lakshminarayanan2017simple}. Financial applications use disagreement across independently trained hedging networks to adjust hedge positions and agreement across LLM agents to filter unreliable information before trading decisions~\citep{poddar2026deephedging,li2026trusttrade}. Unlike sampling-based methods, ensembles vary the model or its parameters rather than repeatedly execute a fixed system. Disagreement is informative only when members are sufficiently diverse; highly correlated members may agree while sharing the same failure. This family can characterize capability uncertainty (\(U_K\)), reliability under distribution shift (\(U_S\)), and uncertainty in model-based financial outcomes (\(U_O\)).

\paragraph{Distributional and statistical uncertainty estimation}
These methods represent uncertainty over a defined target and horizon through predictive distributions, conditional quantiles, intervals, prediction sets, volatility estimates, or tail-risk measures. Established statistical methods such as ARCH/GARCH and quantile regression provide foundations for predictive uncertainty in financial forecasting and risk modeling~\citep{engle1982arch,bollerslev1986garch,koenker1978quantile}. Conformal and evidential methods provide prediction sets, intervals, and uncertainty decompositions~\citep{romano2019conformalized,gibbs2021adaptive,amini2020evidential}. Recent work applies these ideas to time-series foundation models, financial forecasting, and risk estimation~\citep{ansari2024chronos,goel2024var,wong2025volatilityuq,pele2026llmvar}. ProbFM uses evidential uncertainty decomposition for financial forecasting, while Bayesian state filters maintain posterior regime distributions and risk measures in agentic settings~\citep{chinta2026probfm,dixon2026beliefatrisk}. This family primarily characterizes financial outcome uncertainty (\(U_O\)); adaptive procedures may also address reliability under distribution shift (\(U_S\)).

No single representation covers the full framework: evidence verification cannot describe future outcome distributions, while a calibrated distribution does not establish valid evidence or reasoning. Evidence-, model-, and outcome-level representations should remain distinct rather than be collapsed into generic confidence. In agentic workflows, these representations may also be updated or used across dependent stages.

\subsection{Uncertainty Quantification in Financial Agentic Workflows}
\label{subsec:agent_methods}

In financial agentic workflows, local uncertainty estimates may be propagated, aggregated, or used to guide later steps~\citep{zhao2025saup,oh2026agentuq}. Repeated executions of trading-agent systems reveal run-to-run instability~\citep{koviazin2026reproducibility}; agreement across multiple agents provides an ensemble-based reliability measure~\citep{li2026trusttrade}; Bayesian filters update beliefs and risk measures over time~\citep{dixon2026beliefatrisk}; and trajectory-level probes audit failures across agent roles and stages~\citep{chen2026riskfirst}. These approaches combine the five method families above rather than define a separate agent-specific family. The financial studies reviewed here do not yet provide a general method for preserving the meaning of heterogeneous uncertainty representations across a workflow. Verification, abstention, deferral, position adjustment, and escalation govern how an agentic workflow responds to uncertainty.

\subsection{Evaluating Uncertainty Representations}
\label{subsec:evaluation}

An uncertainty representation should be evaluated against the object it characterizes and, when used downstream, the decision or workflow it is intended to support. For example, among comparable forecasts assigned \(90\%\) probability to a positive market reaction, the event should occur approximately \(90\%\) of the time. The estimate should also identify unreliable cases and, when used for decision support, improve decisions after costs and constraints.

We distinguish three complementary evaluation targets. (1) Task performance measures the underlying output against a reference outcome, such as answer correctness, forecast realization, or financial utility. (2) Uncertainty quality measures whether a score, distribution, interval, or prediction set corresponds to observed errors or outcomes; relevant criteria include calibration, error detection, risk--coverage curves, proper scoring rules, coverage, and sharpness~\citep{brier1950verification,guo2017calibration,geifman2017selective,gneiting2007strictly}. (3) Decision value measures whether acting on the representation improves the workflow or financial action after accounting for verification, abstention, implementation costs, and constraints. Strong task performance does not imply reliable uncertainty, and reliable uncertainty does not by itself establish decision value.

The appropriate test depends on the representation. Evidence judgments require labeled supported and unsupported claims, as provided by atomic factuality and financial verification benchmarks~\citep{min2023factscore,zhang2025faith,panda2026finverbench}. Predictive intervals require empirical coverage at useful width, while tail-risk estimates require exceedance and dependence backtests~\citep{christoffersen1998evaluating,kupiec1995techniques}. Evaluation must also preserve temporal order because a representation calibrated in one regime may fail after a shift~\citep{ovadia2019trust}. Recent financial work extends these requirements to time-gated calibration, agent trajectories, financial biases, market frictions, and deployment assumptions~\citep{ghosh2026finbench,chen2026riskfirst,kong2026bias,ye2026alphaillusion}.

A financial UQ study should report the target and horizon, data vintage and leakage controls, construction and evaluation of the uncertainty representation, robustness across time or regimes, and downstream costs and constraints~\citep{jensen2026implementable,ye2026alphaillusion}.

\begin{table*}[t]
\centering
\caption{Financial task domains, representative tasks, dominant uncertainty sources, and relevant method families.}
\label{tab:workflow_uq}
\scriptsize
\setlength{\tabcolsep}{4.2pt}
\renewcommand{\arraystretch}{1.14}
\begin{tabular}{
  @{}
  p{0.175\textwidth}
  p{\dimexpr0.495\textwidth-6\tabcolsep\relax}
  p{0.085\textwidth}
  p{0.245\textwidth}
  @{}
}
\toprule
\raggedright\textbf{Task domain} & \raggedright\textbf{Representative tasks} & \centering\textbf{Dominant sources}\textsuperscript{a} & \raggedright\textbf{Most relevant method families} \tabularnewline
\midrule
\raggedright\textbf{Information and Document Intelligence} & \raggedright Retrieval and RAG; financial QA; information extraction; table, chart, and multimodal document understanding; grounded summarization & \centering\(U_E,U_K,U_R\) & \raggedright Evidence and Provenance Verification \tabularnewline
\midrule
\raggedright\textbf{Analysis, Valuation, and Advisory} & \raggedright Disclosure and event interpretation; financial-statement analysis; valuation; investment research and due diligence; investment and wealth advisory & \centering\(U_R,U_K,U_D\) & \raggedright Direct Uncertainty Estimation; Sampling-Based Consistency Estimation \tabularnewline
\midrule
\raggedright\textbf{Forecasting and Risk Modeling} & \raggedright Return, earnings, cash-flow, and volatility forecasting; credit and default risk; VaR/ES; stress testing & \centering\(U_O,U_S\) & \raggedright Distributional and Statistical Uncertainty Estimation; Ensemble-Based Uncertainty Estimation \tabularnewline
\midrule
\raggedright\textbf{Financial Decision-Making} & \raggedright Security selection and trading; portfolio allocation and rebalancing; hedging and position sizing; execution and market making; lending and underwriting & \centering\(U_D,U_O,U_S\) & \raggedright Distributional and Statistical Uncertainty Estimation; Direct Uncertainty Estimation; Ensemble-Based Uncertainty Estimation \tabularnewline
\midrule
\raggedright\textbf{Accounting, Compliance, and Financial Operations} & \raggedright Accounting and reporting; regulatory interpretation and compliance review; KYC/AML investigation; audit and internal controls; payments, settlement, reconciliation, and claims review & \centering\(U_E,U_R,U_D\) & \raggedright Evidence and Provenance Verification; Direct Uncertainty Estimation \tabularnewline
\bottomrule
\end{tabular}
\par\smallskip
{\footnotesize\raggedright
\textsuperscript{a} \(U_E\): Evidence and Grounding Uncertainty; \(U_S\): Distribution Shift Uncertainty; \(U_K\): Knowledge and Capability Uncertainty; \(U_R\): Reasoning and Interpretation Uncertainty; \(U_O\): Financial Outcome Uncertainty; and \(U_D\): Decision Uncertainty under Constraints.
\par}
\end{table*}

\section{Uncertainty Quantification Across Financial Tasks}
\label{sec:domains_tasks}

Building on Section~\ref{sec:methods}, we examine which uncertainty sources arise in representative financial tasks supported by foundation models and which methods are most relevant. Table~\ref{tab:workflow_uq} summarizes this mapping, which reflects conceptual relevance rather than the frequency of methods in the current literature.

\subsection{Information and Document Intelligence}

Information and Document Intelligence covers retrieval, extraction, financial QA, document understanding, and grounded summarization; economic interpretation belongs to Analysis, Valuation, and Advisory. Reliability depends on whether the output is supported by the correct document, fiscal period, unit, version, and information available at the decision time. Evidence and Provenance Verification is therefore the primary method family. Direct or sampling-based estimates may flag unreliable cases, but they cannot establish that the supporting evidence is valid. Evaluation should examine both output correctness and supporting evidence. FinRAGBench-V and PHANTOM provide evidence- or hallucination-oriented labels, while FinQA and TAT-QA provide correctness labels that can support UQ evaluation~\citep{zhao2025finragbenchv,ji2025phantom,chen2021finqa,zhu2021tatqa}.

\subsection{Analysis, Valuation, and Advisory}

Analysis, Valuation, and Advisory covers financial interpretation, valuation, research, and advice rather than fact retrieval, future-outcome forecasting, or execution. The key question is whether a conclusion follows from the evidence, comparison basis, and financial assumptions. Direct Uncertainty Estimation can estimate confidence or error risk, while Sampling-Based Consistency Estimation can reveal sensitivity to alternative generations or formulations. For valuation, uncertainty should be tied to the assumptions and the resulting range of implied values. In advisory settings, it should support suitability assessment, abstention, or escalation under client-specific objectives and constraints. RealFin and FinBen provide relevant reasoning cases, but dedicated UQ protocols for valuation and personalized advice remain limited~\citep{dai2026realfin,xie2024finben}.

\subsection{Forecasting and Risk Modeling}

Forecasting and risk modeling concern conditional distributions of future returns, earnings, cash flows, volatility, defaults, and losses. The main sources are financial outcome uncertainty and reliability under changing regimes. Distributional and Statistical Uncertainty Estimation is the primary method family because the target is usually a conditional distribution, quantile, interval, or tail-risk measure; ensembles can additionally characterize variation across model realizations. Average forecast error is insufficient: forecasts require calibration, coverage, and sharpness, while risk estimates require tail and stress-period validation. FinBench provides a time-gated calibration setting, complementing established backtests for interval and tail-risk forecasts~\citep{ghosh2026finbench,kupiec1995techniques,christoffersen1998evaluating}.

\subsection{Financial Decision-Making}

Financial decision-making converts forecasts, risk estimates, or analytical conclusions into feasible actions: predicting default is a risk-modeling task, whereas approving a loan is a decision. The uncertainty concerns which feasible action is preferred and how stable that choice is to uncertain outcomes and model inputs. Distributional methods represent outcome uncertainty, while direct and ensemble-based estimates can assess model reliability or support selective action and position adjustment. A decision-level UQ claim should show that uncertainty changes and improves the selected action, not merely that it predicts errors. Comparisons should use matched information, costs, execution timing, constraints, and intervention budgets~\citep{jensen2026implementable,ye2026alphaillusion}.

\begin{figure*}[t]
  \centering
  \includegraphics[width=0.6\textwidth]{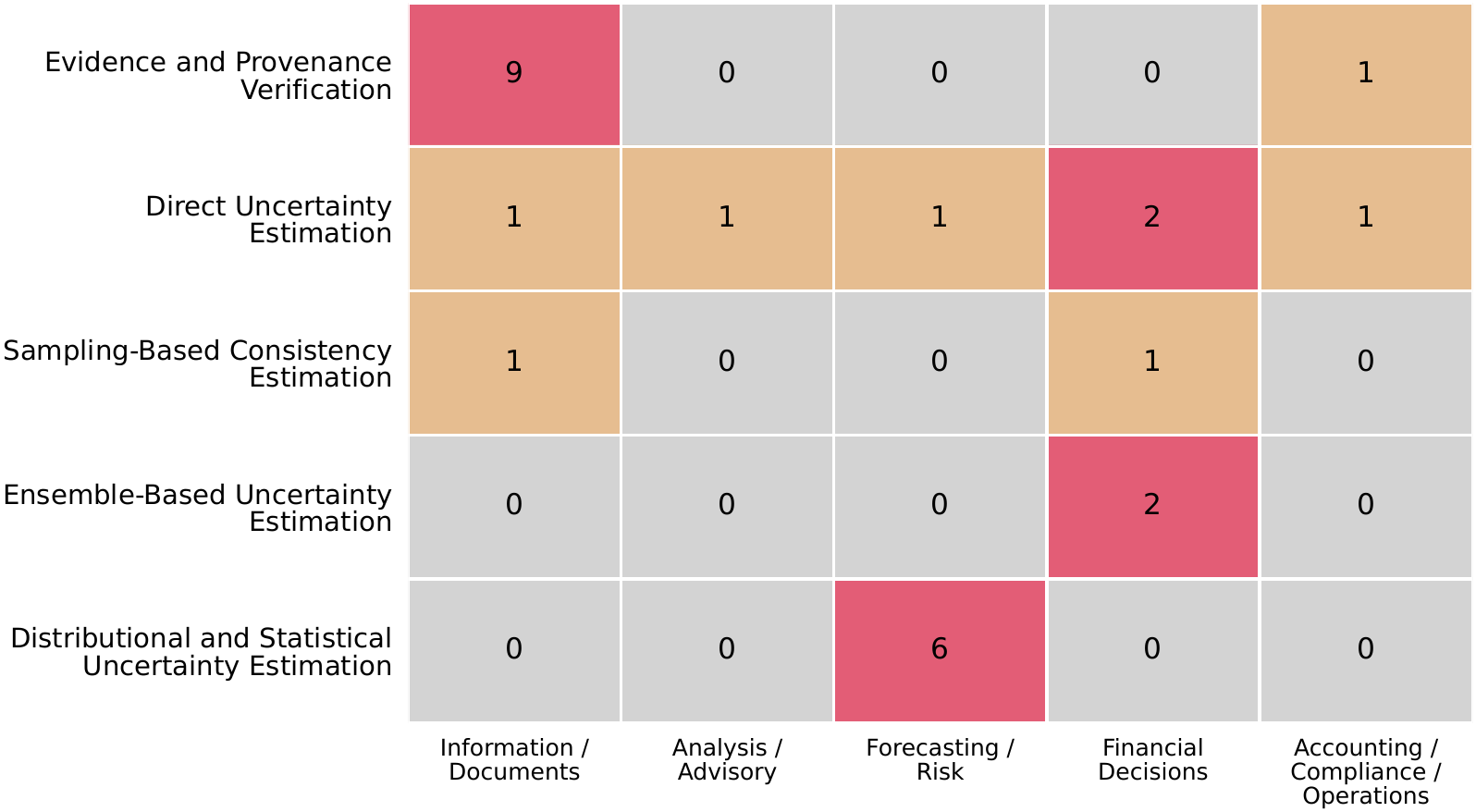}
  \Description{A five-by-five heatmap of unique-paper counts across five uncertainty method families and five financial task domains. Evidence and provenance verification is concentrated in information and document tasks, while distributional and statistical uncertainty estimation is concentrated in forecasting and risk.}
  \caption{Method--task coverage in the included finance-specific studies. Cells report unique-paper counts across 24 papers and 26 assignments; gray denotes zero, orange one, and red more than one study.}
  \label{fig:landscape}
\end{figure*}

\subsection{Accounting, Compliance, and Financial Operations}

Accounting, Compliance, and Financial Operations covers model-supported judgments and exception handling in accounting, regulatory review, KYC/AML, audit, payments, settlement, reconciliation, and claims. A fraud or default probability belongs to Forecasting and Risk Modeling; investigation, rule application, approval, reporting, and escalation belong here. The main uncertainties concern the validity of records and rules, the interpretation of accounting or regulatory requirements, and whether the resulting action is permissible. Evidence and Provenance Verification is central, while Direct Uncertainty Estimation can support selective review and escalation. Evaluation should test whether rules are applied correctly, conclusions can be traced to the relevant records and requirements, and asymmetric errors and regulatory changes are handled appropriately. FAITH, FinVerBench, and FinTrust cover selected aspects of statement verification and trustworthiness, but not the full range of these processes~\citep{zhang2025faith,panda2026finverbench,hu2025fintrust}.

Because an agentic workflow describes how a system executes a financial task rather than a separate task domain, we classify agents by the evaluated output; tool-call validity and trajectory stability remain cross-cutting concerns~\citep{choi2025finagentbench,zhu2026finmcp,chen2026riskfirst}.

\section{Method--Task Landscape and Research Priorities}
\label{sec:landscape}

We code the reviewed finance studies by their primary method family and financial task label(s). Figure~\ref{fig:landscape} reports aggregate method--task coverage; the paper-level coding table is omitted due to space limits.

\paragraph{Selection and coding criteria}
We retain a study only if it satisfies two criteria. First, it must construct, use, or directly evaluate an uncertainty or reliability representation covered by Section~\ref{sec:methods}, such as evidence support, confidence or error estimates, disagreement measures, predictive distributions or sets, risk measures, or uncertainty-conditioned actions. Studies reporting only aggregate task performance, e.g., accuracy, return, or Sharpe ratio, are excluded. Second, it must report substantive quantitative evaluation on financial data or workflows in at least one task domain from Section~\ref{sec:domains_tasks}. General methods without financial experiments, surveys, and conceptual studies are excluded. Each paper receives one primary method family according to the mechanism producing its main uncertainty result. Multiple task labels require distinct empirical results for those domains; passing through several workflow stages alone does not create multiple assignments. Duplicate versions are merged. This yields 24 papers and 26 method--task assignments.

\paragraph{Coverage patterns}
Three patterns stand out in the included corpus. First, method--task specialization is strong: Evidence and Provenance Verification accounts for nine of eleven assignments in Information and Document Intelligence, while all six Distributional and Statistical assignments occur in Forecasting and Risk Modeling. Direct Uncertainty Estimation is the only family represented across all five task domains. Second, task coverage is uneven. Information and Document Intelligence and Forecasting and Risk Modeling contain most assignments, whereas Analysis, Valuation, and Advisory and Accounting, Compliance, and Financial Operations are sparsely represented. Financial Decision-Making is represented mainly by trading, hedging, and agent-based settings, with limited evidence in lending, underwriting, and execution. Third, cross-stage UQ is rare in the included corpus: only two studies span more than one task domain, and we did not identify an included study that carries evidence or reasoning reliability through predictive uncertainty into constrained financial action. Most included work studies uncertainty within parts of the workflow rather than across the path from evidence to action. A zero indicates no included study met our criteria for that method--task pair; it does not imply that no related work exists.

\paragraph{Research priorities}
Three priorities follow from these patterns. First, future work should connect uncertainty across stages while preserving the meaning of evidence, reasoning, outcome, and decision uncertainty rather than collapsing them into a single confidence score. Second, UQ should extend to under-studied settings, including valuation and advisory, accounting and compliance, and financial decisions beyond trading and hedging, with evaluation matched to task-relevant uncertainty. Third, financial decision and agent studies should establish decision value by testing whether uncertainty changes and improves the action under matched information, timing, implementation costs, constraints, and intervention budgets. Retrieval or verification may reduce epistemic uncertainty; when it cannot be resolved before action, abstention, escalation, or position adjustment can limit its consequences.

\section{Conclusion}

This survey develops a source-aware framework for uncertainty in financial foundation-model workflows, distinguishing what remains unresolved across evidence and context, model and inference, and outcome and action. It organizes UQ methods by how they construct or verify uncertainty and reliability representations, then examines their applicability across representative financial tasks. The resulting method--task landscape shows how current empirical work is distributed and supports research priorities for connecting uncertainty across stages, extending it to under-studied financial settings, and establishing its value for downstream decisions.

The review is bounded by its focus on foundation-model and agentic financial workflows and by the databases, search terms, and literature cutoff described in Section~\ref{sec:intro}; it does not attempt to cover the full literature on classical financial uncertainty modeling. The field is also developing quickly, and several recent studies remain preprints whose findings may change with broader evaluation or subsequent publication.

Taken together, the framework and empirical synthesis shift the focus from confidence in a final output to the different forms of uncertainty that arise across a financial workflow. The literature remains largely separated across tasks and stages, with limited evidence connecting evidence and reasoning reliability to predictive uncertainty and constrained action. This survey provides a unified foundation for understanding, evaluating, and using uncertainty in financial foundation models.

\bibliographystyle{ACM-Reference-Format}
\bibliography{references}

\end{document}